# Ultrafast X-ray interaction with photovoltaic materials: Thermal and nonthermal responses

Aldo Artímez Peña[1,2], Nikita Medvedev[1,3]

1) Institute of Physics, Czech Academy of Sciences, Na Slovance 1999/2, 182 00 Prague 8, Czech Republic

2) Higher Institute of Technologies and Applied Sciences, University of Havana, Ave. Salvador Allende 1110, 10 400, La Habana, Cuba

3) Institute of Plasma Physics, Czech Academy of Sciences, Za Slovankou 3, 182 00 Prague 8, Czech Republic

## Abstract

Cadmium telluride (CdTe), lead sulfide (PbS), and indium tin oxide (ITO) play crucial roles in various electronic applications where laser treatment enables precise modification of their distinctive electronic characteristics. This study utilizes the XTANT-3 hybrid/multiscale model to investigate the microscopic response of these materials to ultrafast X-ray irradiation. The model simultaneously traces intertwined processes of non-equilibrium dynamics of both electrons and atoms, nonadiabatic coupling, nonthermal melting, and bond breaking due to electronic excitation. Among the materials studied, CdTe exhibits the highest radiation resistance, similar to CdS. At the respective threshold doses, the melting is primarily thermal, driven by electron-phonon coupling, which is accompanied by the band gap closure. Additionally, all materials exhibit nonthermal melting at higher doses. When accounting for energy dissipation pathways and material recrystallization processes, damage thresholds increase substantially. In CdTe and PbS, below 1.5 eV/atom, the band gap returns to its original value upon recrystallization. As the dose increases, the resulting cooled material becomes increasingly amorphous, progressively reducing the band gap until a stable configuration is reached. Notably, in a narrow window of deposited doses, ITO exhibits transient superionic behavior, with the liquid oxygen but solid In and Sn sublattices. At 0.6 eV/atom in CdTe and 0.4 eV/atom in PbS and ITO, material ablation from the surface occurs. These findings indicate that femtosecond laser technology offers promising opportunities for precise band gap engineering in various photovoltaic semiconductor devices.

# I. Introduction

Modern electronic device manufacturing involves a variety of materials. Key semiconductor compounds nowadays include cadmium telluride (CdTe), lead sulfide (PbS), and indium tin oxide (ITO), which have gained prominence across multiple applications, such as photovoltaic technologies[1–3], light-emitting diodes[4,5], and radiation detectors in Free Electron Laser (FEL) sources[6]. Zincblende CdTe, the most common phase of this compound, has a direct bandgap of 1.45 eV[7] while PbS is a narrow band gap material (0.41 eV), making it sensitive to infrared light, with thin-film configurations displaying quantum confinement effects[8]. ITO is a tin-doped $In_2O_3$-based n-type wide-bandgap semiconductor (4.0 eV) with Sn dopant levels forming below the bottom of the conduction band, making it a nearly transparent conducting material.[1–3]

Laser-based processing is a fundamental approach for semiconductor engineering, , machining, and nano-patterning [9,10]. Laser irradiation induces spatially confined phase transitions in target materials, enabling precise property modification unachievable by any other means. Melting and ablation thresholds under nanosecond-pulse irradiation have been previously documented for CdTe[11,12]. However, comparable data for PbS or ITO remain unreported in the current literature. Furthermore, high-dose-rate irradiation (ultrashort intense pulses) may trigger alternative kinetic pathways and produce damage distinct from its low-dose-rate counterpart[13].

Free-electron lasers produce intense femtosecond pulses of extreme ultraviolet (XUV)/X-ray radiation.[13–16] Due to the ultrashort pulse duration, such irradiation achieves extremely high dose rates. It enables the generation and examination of highly nonequilibrium states of matter under extreme conditions[13,17] and allows for unprecedented control of the material modifications [18]. It is, therefore, a promising tool for materials processing.[19,20]

Laser interaction with matter involves several distinct stages[21,22]: Initially, photon absorption by electrons occurs, promoting electrons to high-energy levels in the material. This includes valence-to-conduction band transitions in semiconducting materials. When the photon energy is sufficiently high, i.e., in the case of extreme ultraviolet (XUV) or X-ray lasers, absorption primarily takes place in core atomic shells, exciting bound electrons to unoccupied states and creating core-level holes. These core holes undergo Auger (or radiative, for heavy element deep shells) decay, typically at femtosecond timescales [23,24].

Subsequently, excited electrons scatter with the surrounding matter through various mechanisms: generating additional excited electrons (impact ionization), with the collective electron modes (plasmons), and with atoms and their collective modes (phonons), transferring energy to the atomic lattice [22,24]. All these processes ultimately lead to equilibration of the electronic ensemble, reaching a Fermi-Dirac distribution at sub-picosecond timescales.

Energy transfers from electrons to atoms through two main pathways: nonadiabatic electron-ion (electron-phonon) coupling, occurring at picosecond timescales; and adiabatic modification of the interatomic potential. At high radiation doses, the latter mechanism may trigger nonthermal melting or bond breaking, causing ultrafast atomic disordering even in the absence of significant thermal heating [25,26]. Above non-thermal melting thresholds, electron-driven modifications of the interatomic potential may accelerate atoms and heat the lattice at sub-picosecond scales[27]. Combined atomic heating and bond disruption may induce phase transitions, producing novel material states including alternative solid or liquid phases, or even transient unusual states outside of the equilibrium phase diagram [28,29].

This work aims to examine the processes triggered in CdTe, PbS, and ITO by ultrafast intense XUV/X-ray irradiation, determining the respective damage thresholds and mechanisms, along with those states produced as part of phase transitions.

## II.  Model

Damage kinetics in CdTe, PbS, and ITO induced by ultrafast X-ray or XUV radiation are simulated with the hybrid (multiscale) code XTANT-3[30]. The code unifies multiple theoretical models describing various processes mentioned above [31]. A comprehensive description of the models and their computational implementation are available, e.g., in Ref. [32]; below, a condensed overview of the physics and the methodology of their numerical description, are presented.

The X-ray/XUV photon absorption, subsequent electron cascades, and core-hole Auger relaxation events, are modelled with event-by-event (analog) transport Monte-Carlo (MC) simulations [31,33,34]. Data on photoabsorption cross sections, Auger decay times, and ionization potentials of core shells are sourced from the EPICS2023 database [35]. Electron kinetics within the MC module continues until the kinetic energy of a particle decreases to the chosen cutoff of 10 eV, counted from the bottom of the conduction band. Modelling of fast electron elastic collisions relies on the screened Rutherford cross-section with the modified Molier screening parameter [34]. The Ritchie-Howie complex-dielectric-function (CDF) formalism is used to describe the inelastic scattering (impact ionization of core holes and valence band and scattering on plasmons) [36]. Material-specific CDF parameters are determined using the single-pole approximation [37]. Statistical reliability of the MC simulations is ensured by averaging over 50,000 iterations[31,38].

Electrons with energies below the cutoff, populating the evolving valence and conduction bands, are traced with the distribution function evolving via the Boltzmann collision integrals (BCI). The current implementation assumes adherence to the Fermi-Dirac distribution (instantaneous electron thermalization approximation in electron-electron scattering) [39]. The matrix element for the nonadiabatic energy exchange between these electrons and atoms (electron-phonon coupling) is derived from the transient tight binding (TB) Hamiltonian with the dynamical coupling method [40].

The transient electronic orbitals (energy levels, band structure) are evaluated through the transferable tight binding method. The same approach is used in the calculation of the interatomic forces [41]. The transient Hamiltonian, dependent on the spatial coordinates of all the atoms within the simulation box, is diagonalized at each timestep of the simulation, tracing the evolution of the electronic states and the atomic potential energy surface as the system responds to excitation. For each material, we employ the periodic table baseline parameters (PTBP)[42,43], which use an $sp^3d^5$ basis set for the linear combination of atomic orbitals within the DFTB framework.

Atomic motion is traced with classical Molecular Dynamics (MD) simulations. The interatomic forces are derived from the TB Hamiltonian and the transient electron distribution functions (fractional electronic populations traced with the BCI method above). This approach captures the modifications in the interatomic potential arising from alterations in the electronic distribution due to X-ray-pulse excitation and high-energy electron scattering events. This way, the model is capable of reproducing the above-mentioned nonthermal phase transitions[26,44]. Non-adiabatic (electron-phonon) energy transfer, computed through BCI methodology, is delivered to atomic ensemble *via* velocity scaling algorithms applied every timestep during the simulation[40].

The propagation of atomic trajectories uses Martyna-Tuckerman 4$^{th}$ order algorithm with a timestep of 1 fs [45]. CdTe and PbS simulations utilize 216-atom supercells. Unit cell atomic coordinates are obtained from Ref. [46]. ITO supercell contains 320 atoms and is set by randomly replacing 10% of the In atoms with Sn atoms in the $In_2O_3$ structure, also taken from Ref. [46]. These supercell sizes are sufficient for reliable simulations[33]. Periodic boundary conditions are employed to simulate the relevant materials in the bulk.

The methods for the calculation of electronic heat capacity and heat conductivity are detailed in Refs. [32] and [47], using 7x7x7 k-point grid. Dynamical coupling formalism is used for the evaluation of the electron-ion coupling parameter calculations, averaged over up to 100 independent realizations[40].

Simulations start 200 fs prior FEL pulse arrival, allowing for the atomic equilibration, and continue for 15 ps post-irradiation for ITO and 30 ps for CdTe and PbS supercells (with a gaussian laser pulse centered at 0 fs) [48]. Long-term effects of irradiation (section III.C) are modelled using Berendsen thermostat set at room temperature with 1 ps (1000 fs) characteristic cooling time[32]. Simulations of thin layers (section III.D) include periodic boundaries along X and Y axis, and free surfaces along Z. Atomic snapshots are visualized with the help of OVITO [49].

XTANT-3 was previously validated against experimental data for damage kinetics in various irradiated materials, showing a reasonable agreement (see, e.g., Refs. [29,33,44,50]).

## III. Results

### A. Thermodynamic properties

Electron heat capacity, electron heat conductivity, and electron-phonon coupling parameter in CdTe, PbS, ITO (and CdS and pure $In_2O_3$ for comparison) are calculated with XTANT-3 (Figure 1-Figure 3). These are key parameters in thermodynamic modeling of laser irradiation, such as the two-temperature model and its derivatives [51–53].

As typical for semiconductors [51], the electronic heat capacity is near zero at the electron temperatures below the values comparable with the bandgap, see Figure 1. Above $T_e$~4,000 K, the heat capacity rises sharply in all the materials studied.

The materials under study exhibit low electronic heat conductivity compared to other semiconductors [47]. The maximum values of this parameter for PbS and ITO are within the range typically observed for metals; moreover, the electronic heat conductivity as a function of temperature is similar to that in elemental Pb[47]. This parameter is significantly lower in CdS and CdTe and almost independent of the temperature at values above 15,000 K (Figure 2).

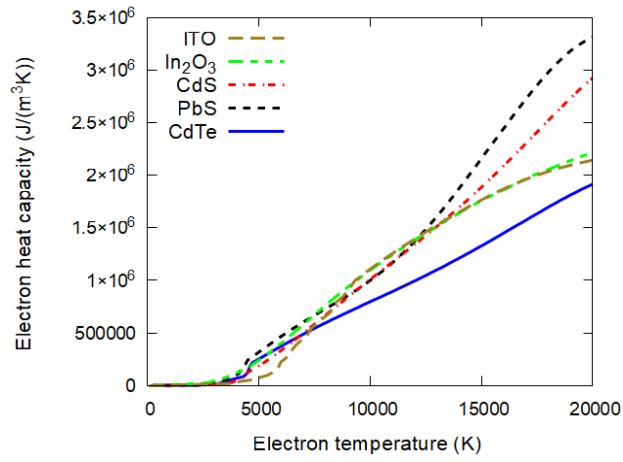

*Figure 1. Electron heat capacity in CdTe, PbS, ITO, and CdS and In$_2$O$_3$ for comparison, calculated with XTANT-3.*

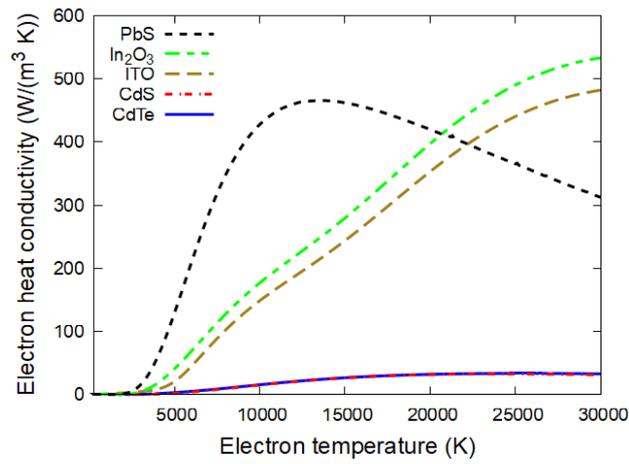

*Figure 2. Electron heat conductivity in CdTe, PbS, ITO, and CdS and In$_2$O$_3$ for comparison, calculated with XTANT-3.*

As shown in Figure 3, the electron-phonon coupling is strongest in ITO, in line with the previous observation that lighter elements typically couple to electrons more efficiently than heavier ones (e.g., compare CdS with CdTe) [40, 51].

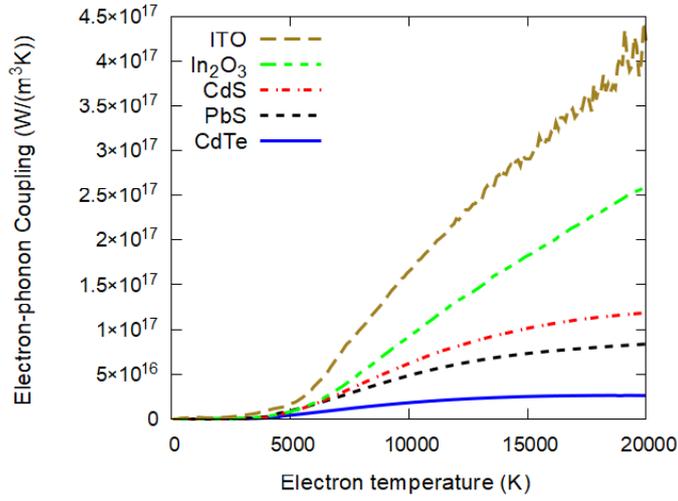

*Figure 3. Electron-phonon (electron-ion) coupling in CdTe, PbS, ITO, and CdS and In$_2$O$_3$ for comparison, calculated with XTANT-3.*

## B. Ultrafast damage in the bulk

A sequence of simulations was performed, varying the irradiation dose to find the phase transition thresholds. The atomic snapshots in Figure 4-Figure 6 show the material response to below and above the threshold doses. CdTe disorders at the dose of ~0.4-0.5 eV/atom (Figure 4), while PbS and ITO do so at ~0.2-0.3 eV/atom (Figure 5) and ~0.3-0.4 eV/atom (Figure 6), respectively. For each material, at doses above the corresponding threshold, the atomic lattice loses stability and turns into a disordered liquid-like state.

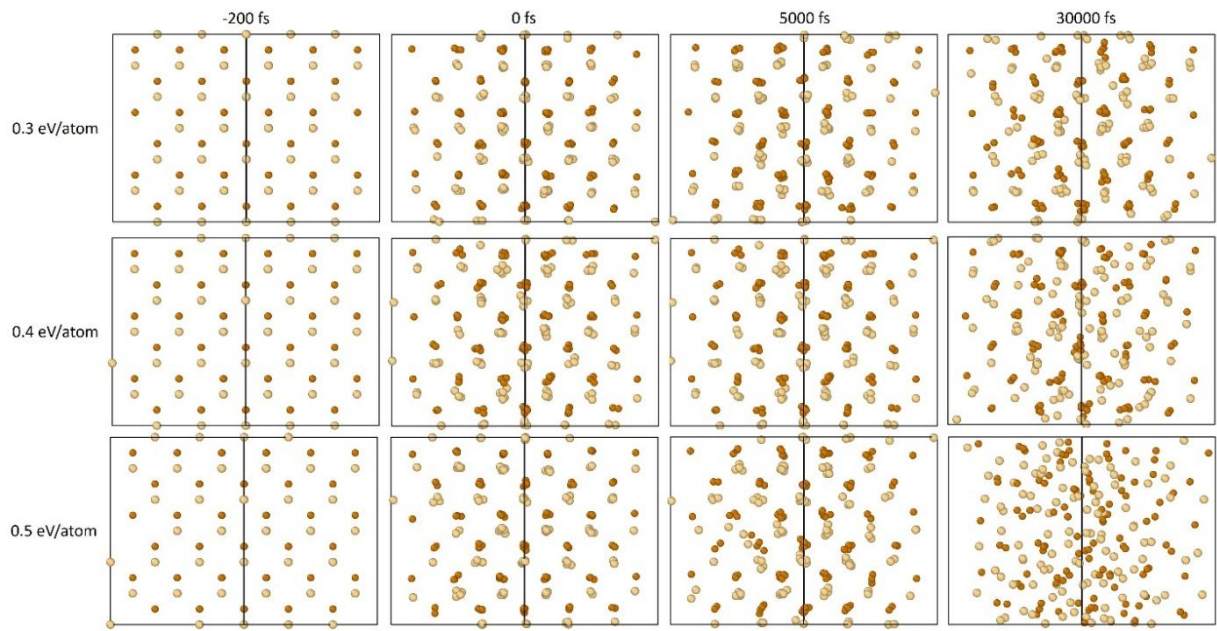

*Figure 4.* *Atomic snapshots of CdTe in the zincblende structure irradiated with different doses. Brown balls are Te; light pink balls are Cd.*

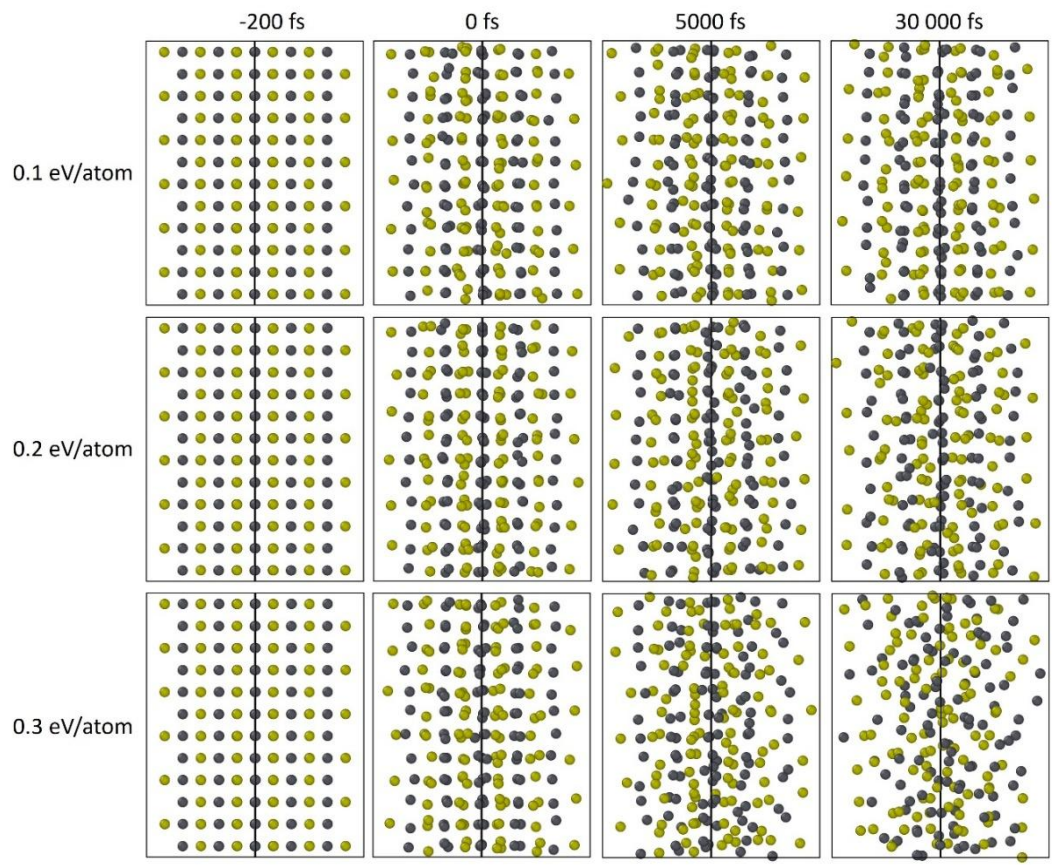

*Figure 5.* *Atomic snapshots of PbS irradiated with different doses. Grey balls are Pb; yellow balls are S.*

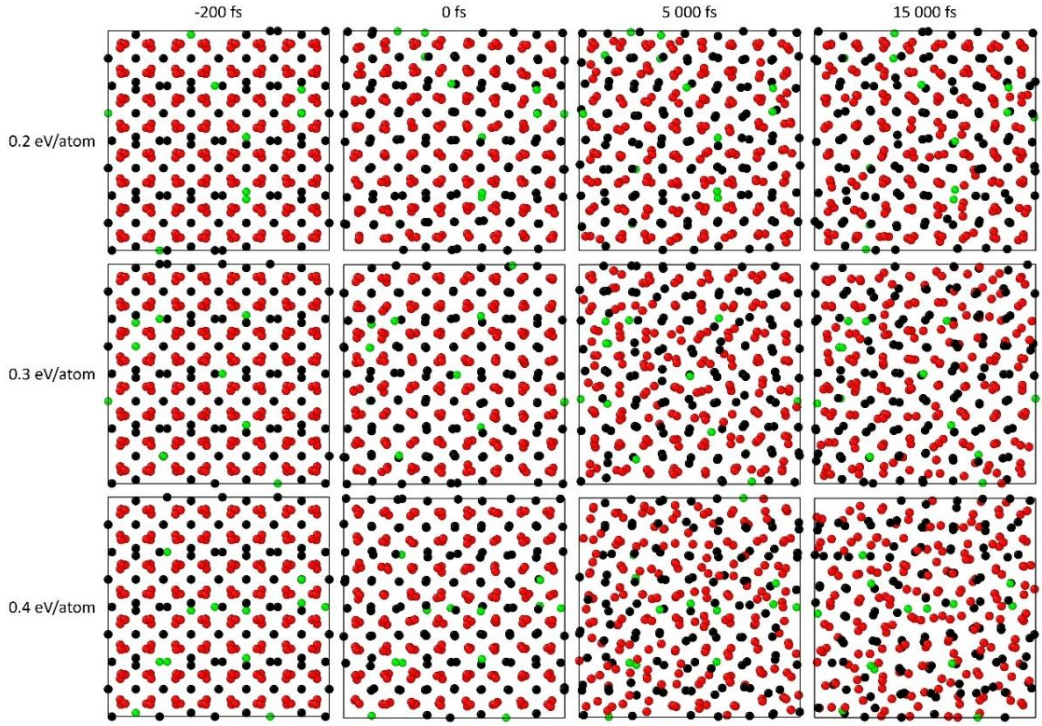

*Figure 6. Atomic snapshots of ITO irradiated with different doses. Black balls are In; green balls are Sn; red balls are O.*

The melting observed at the near-threshold doses is thermal, induced by atomic heating via electron-phonon coupling. This can be established by a comparison with the Born-Oppenheimer (BO) simulation, which excludes the electron-phonon coupling, and thus nonadiabatic heating of the atomic system[54]. The BO simulations show that the nonthermal damage onsets at higher doses for each material (**Table 1**).

*Table 1. Thermal and nonthermal phase transition threshold doses in bulk CdTe, PbS, and ITO calculated with XTANT-3*

| Material | Calculated threshold dose (eV/atom) | |
|---|---|---|
| | non-BO simulations (thermal melting) | BO simulations (nonthermal melting) |
| CdTe | 0.4-0.5 | 1.0 |
| PbS | 0.2-0.3 | 0.9 |
| ITO | 0.3-0.4 | 0.8 |

The atomic heating via electron-phonon coupling in non-BO simulations varies with the material; see the example of the equilibration of the electronic and atomic temperatures in Figure 7. In ITO, the coupling parameter reaches its peak of ~$2.5 \times 10^{17}$ W/(m$^3$K) at ~1 ps after the pulse, when the electronic temperature is still relatively high, and the atomic temperature is also close to its maximum[55], while in PbS and CdTe, the coupling parameter reaches its maximum around 4 ps post-irradiation. Afterwards, the coupling parameter decreases with a decrease in the electronic temperature[51].

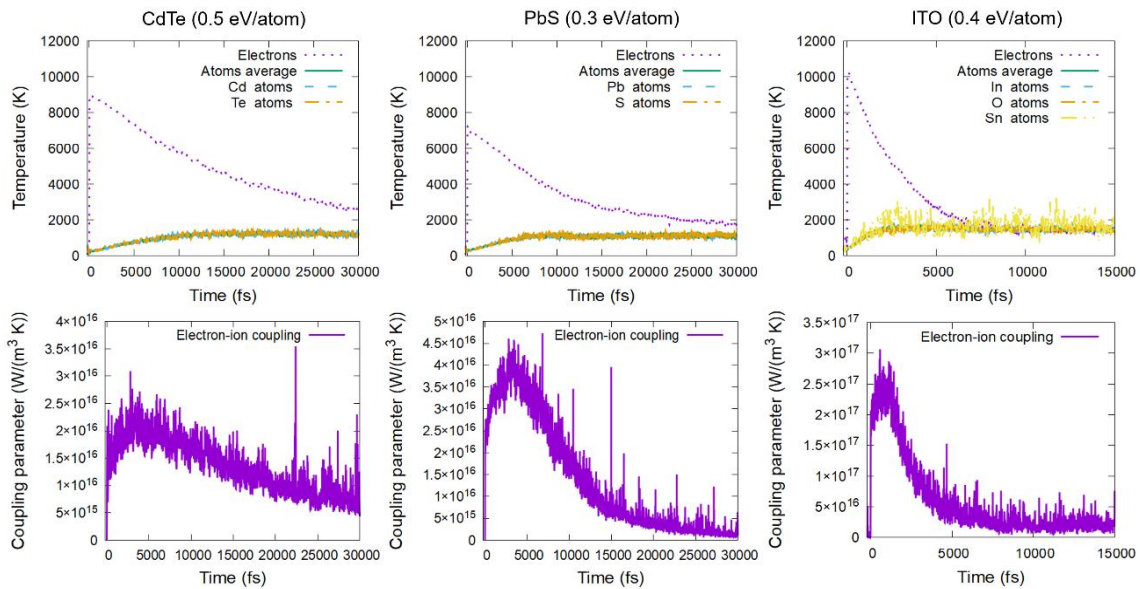

*Figure 7. Electronic and atomic temperatures (top panel) and time dependent electron-ion coupling parameter (bottom panel) in CdTe, PbS, and ITO at the respective doses for phase transition.*

In summary, CdTe appears to be more resistant to ultrafast irradiation than the other two materials under study, and comparable to CdS[56], but with slower phase transition dynamics, which is consistent with Te being heavier than S.

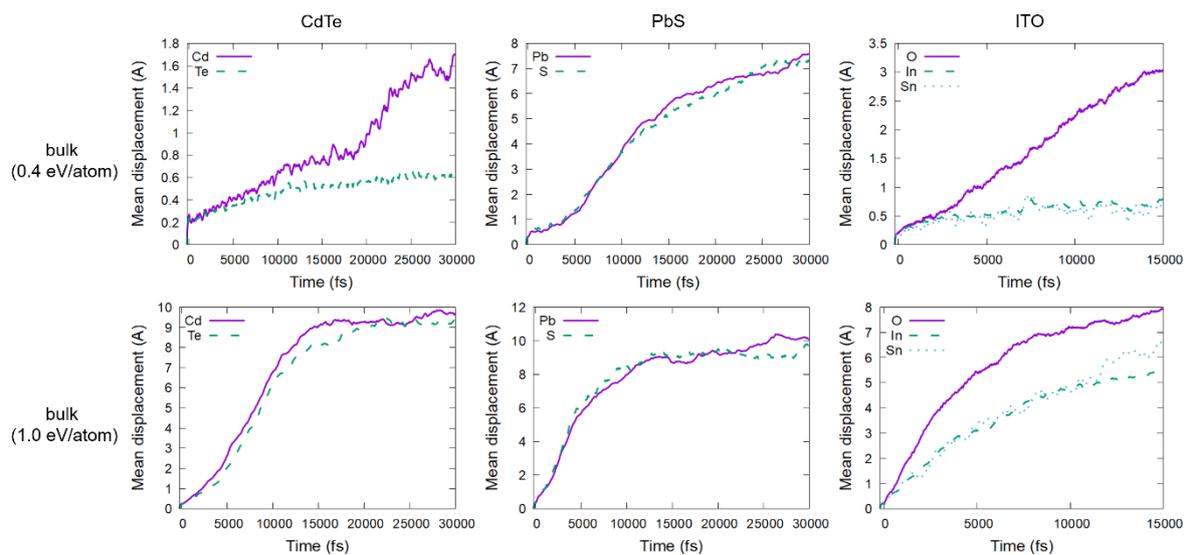

*Figure 8. Mean displacement of each specie in bulk CdTe, PbS and ITO.*

As shown in Figure 8, in CdTe, at a dose of 0.4 eV/atom, the Te mean displacement saturates at ~ 0.6 Å, while the Cd atoms continue to move, demonstrating a diffusive (liquid-like) behavior. A similar behavior is observed in ITO (displacement of In and Sn atoms tends to plateau while it keeps increasing for O) at deposited doses between 0.3 eV/atom and 0.6 eV/atom. This behavior is characteristic of transient superionic states – materials simultaneously exhibiting one solid and another liquid sublattice[28,29 57].

It is interesting to note that the superionic-like state in CdTe (in contrast to $In_2O_3$ and ITO) occurs at the time of ~20 ps, where the electronic and atomic temperature are almost equilibrated (cf. Figure 7), suggesting that the formation of this state is thermal, not triggered by the changes in the interatomic potential induced by high electronic temperatures.

As the irradiation dose increases, the mean displacement of all species tends to equal, indicating that the materials reach complete melting in all sublattices.

In response to irradiation, the band gap in CdTe and PbS shrinks with an increase in the dose (Figure 9). As is typical for ionic materials, the threshold dose for atomic disorder is lower than the threshold dose for the complete band gap collapse[58]. Interestingly, despite the previously discussed higher radiation resistance of CdTe, the dependence of bandgap shrinkage with radiation dose is comparable in both materials (Figure 9): at doses around 0.4 eV/atom, the band gaps of both CdTe and PbS transiently contract to approximately 1 eV. A complete band gap collapse requires doses between 0.6 and 0.7 eV/atom. Thus, XTANT-3 calculations predict that, similar to CdS[56], these materials may transiently form semiconducting or metallic states, depending on the dose. Particularly in CdTe, the results suggest that the band gap may transiently be tuned even at doses below the phase transition threshold.

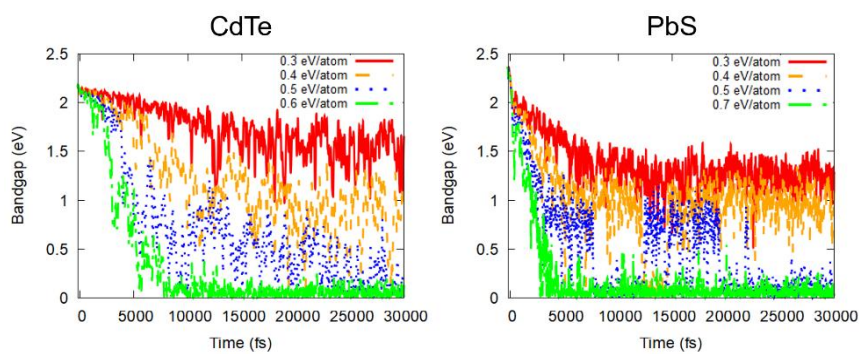

*Figure 9. Band gap of CdTe and PbS irradiated with various doses.*

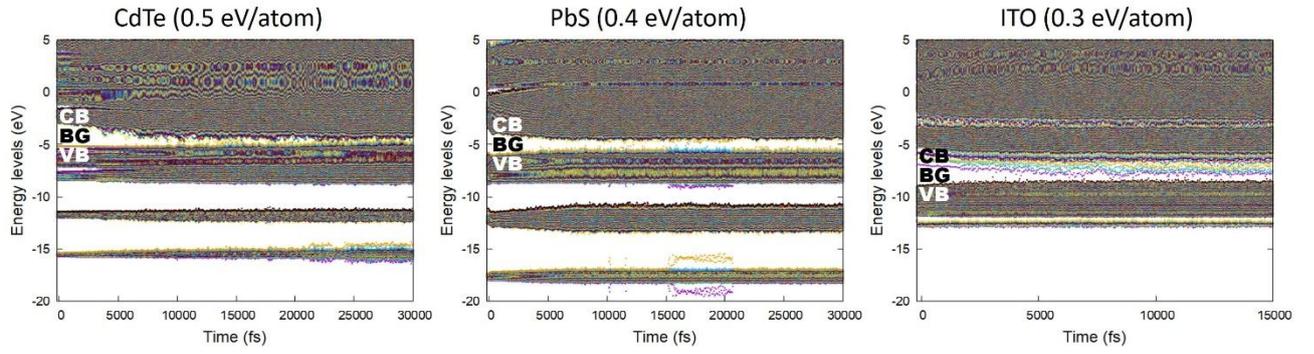

*Figure 10. Electronic energy levels (molecular orbitals, band structure) in CdTe, PbS, and ITO irradiated with different doses. The valence band (VB), the band gap (BG), and the conduction band (CB), are marked.*

Figure 10 shows that, as in CdS[56], the band gap collapse in CdTe takes place mainly *via* lowering of the conduction band. This indicates that the electrons in the conduction band merge with the valence band holes due to energy levels shifting and lose their energy, consequently instigating the nonthermal acceleration of the atoms [27]. This does not seem to be the case in PbS, where the band gap shrinks due to the widening of both the conduction and valence bands, similar to irradiated diamond[33]. In ITO, lowering of dopant levels below the bottom of the conduction band starts at doses around 0.2 eV/atom, while above 0.5 eV/atom, the band gap fully collapses.

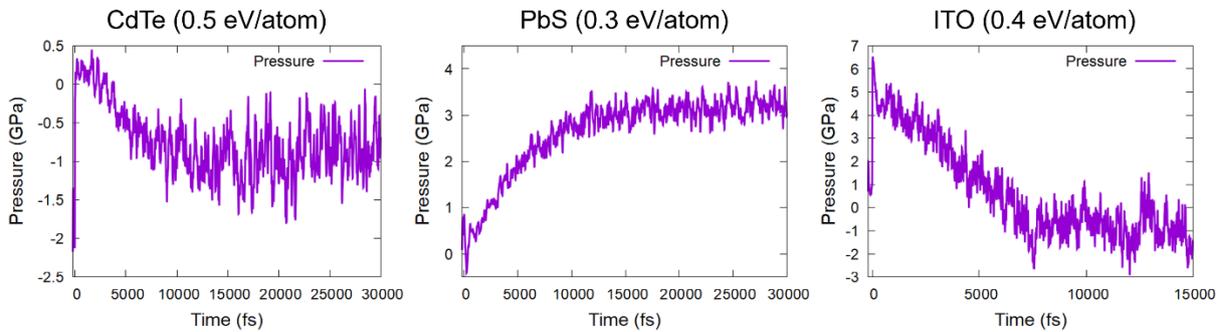

*Figure 11. Pressure in CdTe, PbS and ITO at the respective doses for melting.*

Transition to the disordered state in CdTe and ITO is accompanied by the pressure turning negative, see Figure 11. This indicates that the density of these materials in liquid states is higher than that of the respective crystalline states. Similar formation of a high-density liquid state after irradiation was predicted in silicon and CdS[56]. In contrast, PbS is expected to expand with melting.

## C. Long-time relaxation in the bulk

To estimate the stability of the predicted states and the effects of possible damage recovery in the studied materials, we simulated deposited doses up to 6 eV/atom, allowing for material cooling to the room temperature via a Berendsen thermostat (characteristic cooling time of 1 ps). By the end of these simulations, the atomic temperatures reach room temperature (Figure 12). With equilibration, the mean displacements of the species in the three materials saturate (Figure 13). This suggests that the superionic behavior observed in CdTe and ITO (see section B) is a transient state, and materials resolidify upon cooling.

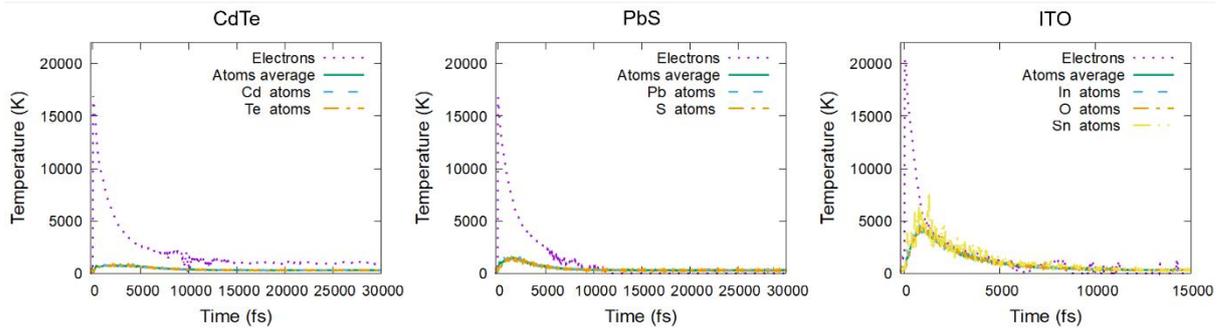

*Figure 12. Electronic and atomic temperatures in CdTe, PbS, and ITO irradiated with 2.5 eV per atom dose, cooled down via thermostat with a characteristic time of 1 ps.*

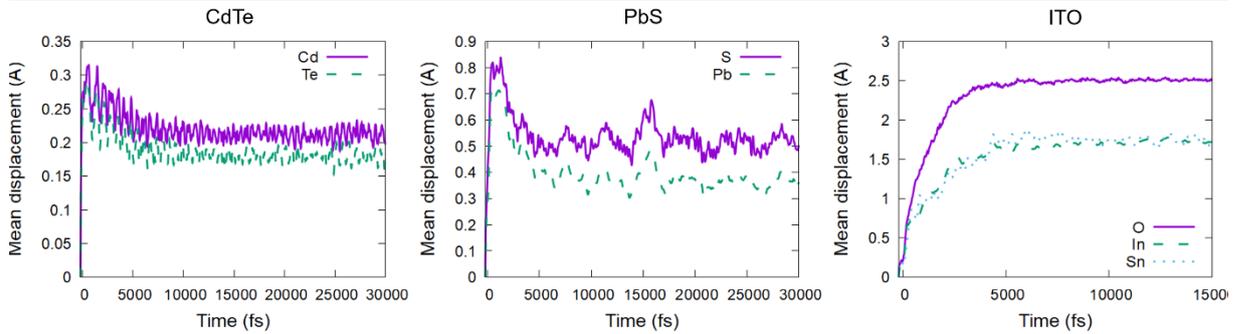

*Figure 13. Mean displacement of each species in the bulk CdTe, PbS, and ITO irradiated with 1 eV/atom deposited dose and cooled down via thermostat with a characteristic time of 1 ps.*

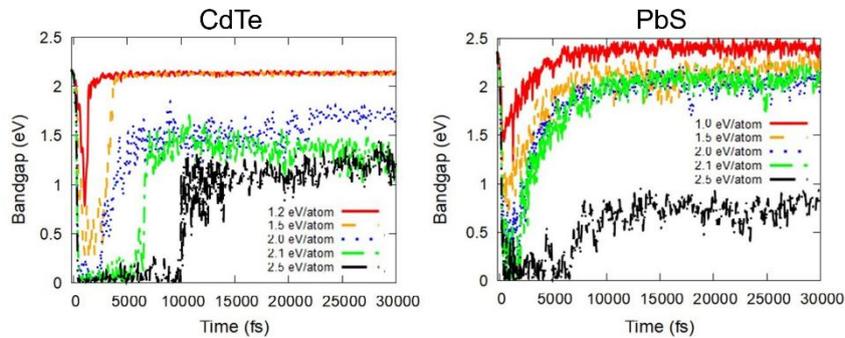

*Figure 14. Band gap values in CdTe, PbS irradiated with various doses, cooled down via thermostat with the characteristic time of 1 ps.*

Accounting for material cooling increases the damage threshold doses. In CdTe, at irradiation doses up to ~1.9 eV/atom, the band gap first collapses, as discussed in the previous section, but opens again with material cooling and recrystallization, returning to the original value. The same behavior is observed in PbS at doses below ~1.0 eV/atom. In both materials the band gap reaches an equilibrium value ($E_{gap}$ ~1.0 eV in CdTe and $E_{gap}$ ~0.8 eV in PbS) at all doses above the threshold of 2.1 eV/atom in CdTe and 2.2 eV/atom in PbS. In a narrow region of doses, the created state has band gap values in between those of the crystalline and stable amorphous phases (Figure 14). This suggests that X-ray ultrashort pulses may be used in material processing to tune the band gap to some degree by tailoring the radiation parameters of the laser pulse.

In ITO, based on the band structure evolution, the band gap recovers to some extent up to doses around 2.2 eV/atom, but the material remains metallic after equilibration at all doses above this value (Figure 15).

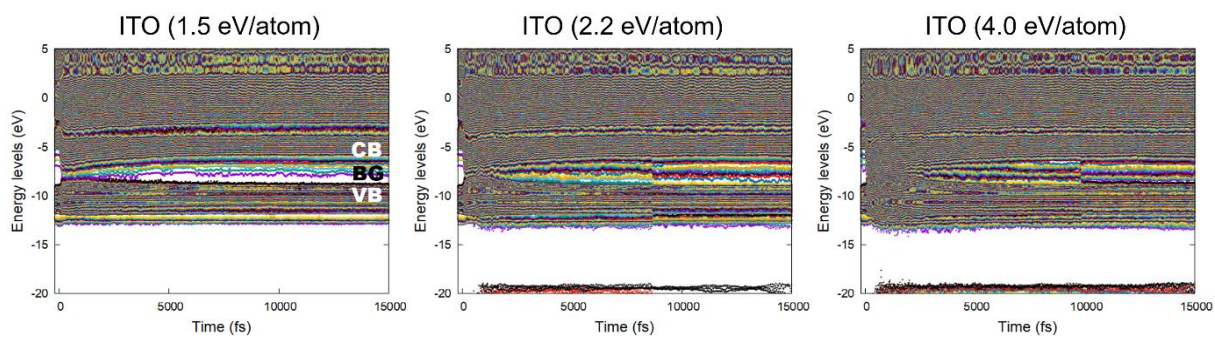

*Figure 15. Electronic energy levels (molecular orbitals, band structure) in ITO irradiated with different doses. The valence band (VB), the band gap (BG), and the conduction band (CB), are marked on the left panel.*

To access the evolving atomic structure in experiments, ultrafast photon or electron diffraction is often employed. Calculated powder diffraction patterns of CdTe irradiated with the dose of 1.0 eV/atom for the probe photon wavelength of 1.54 Å (Figure 16) show that by 1 ps, most of the crystalline peaks are no longer observable but return almost completely as the material cools down and relaxes to equilibrium, suggesting a high degree of recrystallization and confirming the high radiation-resistance of the material.

In ITO and PbS, even though at low doses, only the diffraction peaks at small angles return with relaxation, while the long-range order is lost – the materials do not recover to the extent observed in CdTe, likely due to the presence of light elements in the lattice.

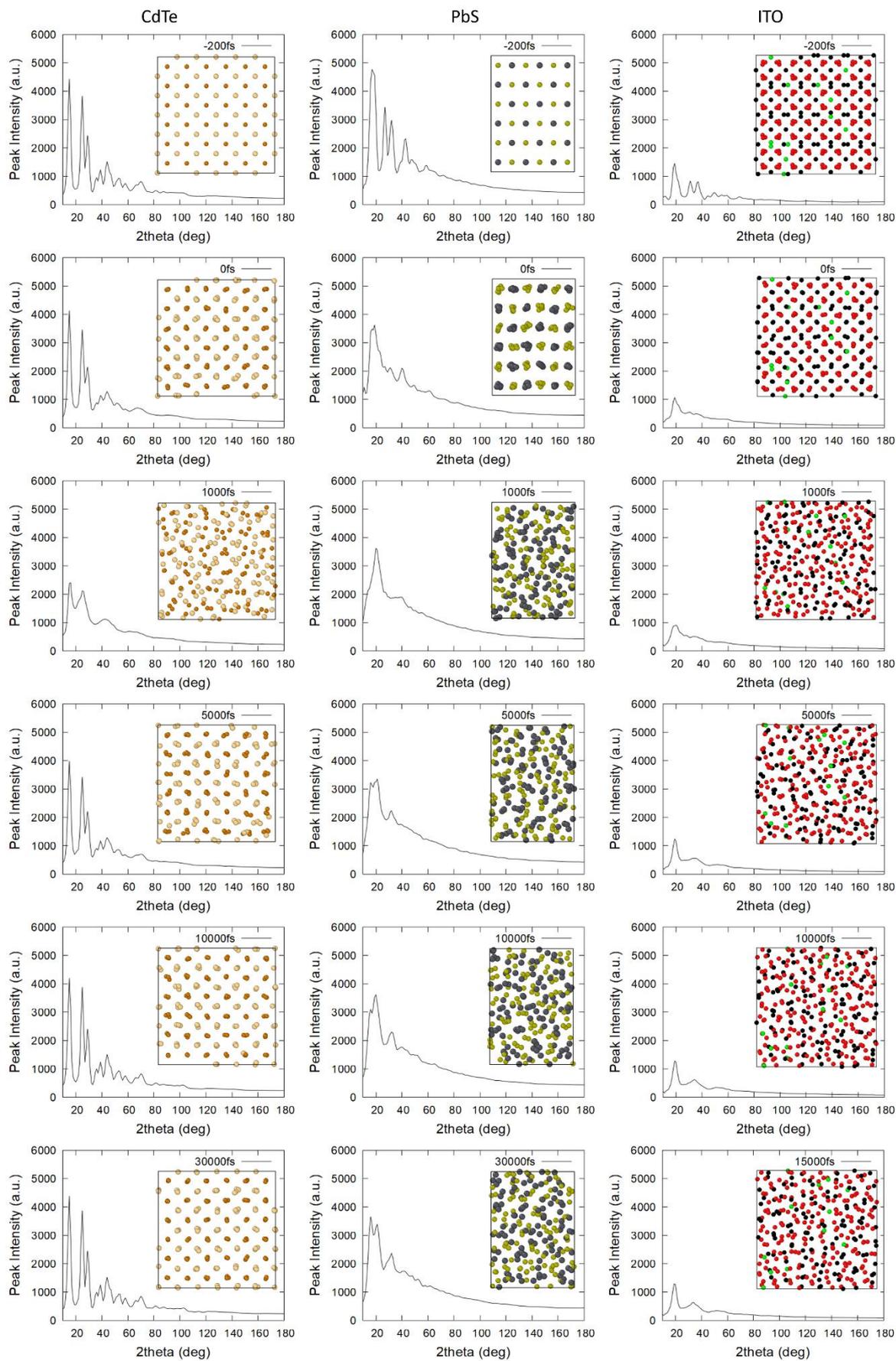

*Figure 16. Powder diffraction patterns (probe wavelength of 1.54 Å) in CdTe, PbS and ITO irradiated with 1.0 eV/atom dose, cooled down with the characteristic time of 1 ps. The insets show the corresponding atomic snapshots.*

### D. Thin layer and surfaces

We have additionally modelled the effects of irradiation on thin layers of the relevant materials. The threshold dose for phase transition is lower for each material in this form (Table 2) compared to that in the bulk (Table 1). This is a consequence of the thin layer expansion, which destabilizes the atomic lattice, thereby lowering the damage threshold.

At doses above the ablation threshold (Table 2) S and O, the most volatile species, are emitted from PbS and ITO surfaces, respectively. Also, in ITO, Sn-O aggregates are emitted (see Figure 17-Figure 19). Interestingly, even though cadmium diffuses more readily within the bulk CdTe (cf. Figure 8), tellurium, the heavier element, is preferentially emitted from the surface (Figure 17, Figure 21), suggesting its lower surface tension.

*Table 2. Melting and ablation threshold doses in CdTe, PbS, and ITO thin layers calculated with XTANT-3*

| Material | Phase transition Threshold (eV/atom) | Ablation Threshold (eV/atom) |
|---|---|---|
| CdTe | 0.3 | 0.6 |
| PbS | 0.2 | 0.4 |
| ITO | 0.2 | 0.4 |

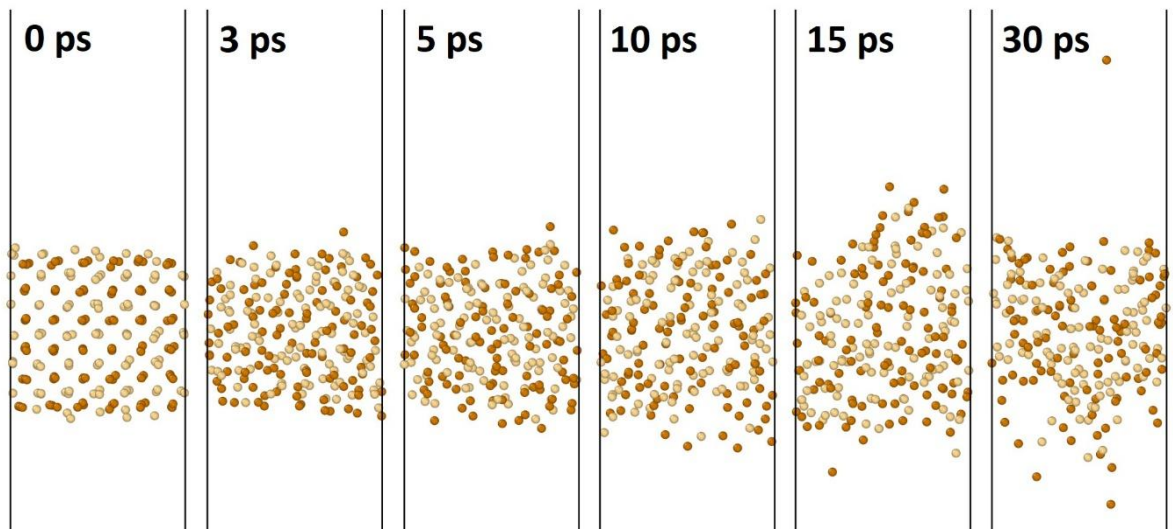

*Figure 17. Atomic snapshots of CdTe thin layer irradiated with 1.0 eV/atom dose. Brown balls are Te; light pink balls are Cd.*

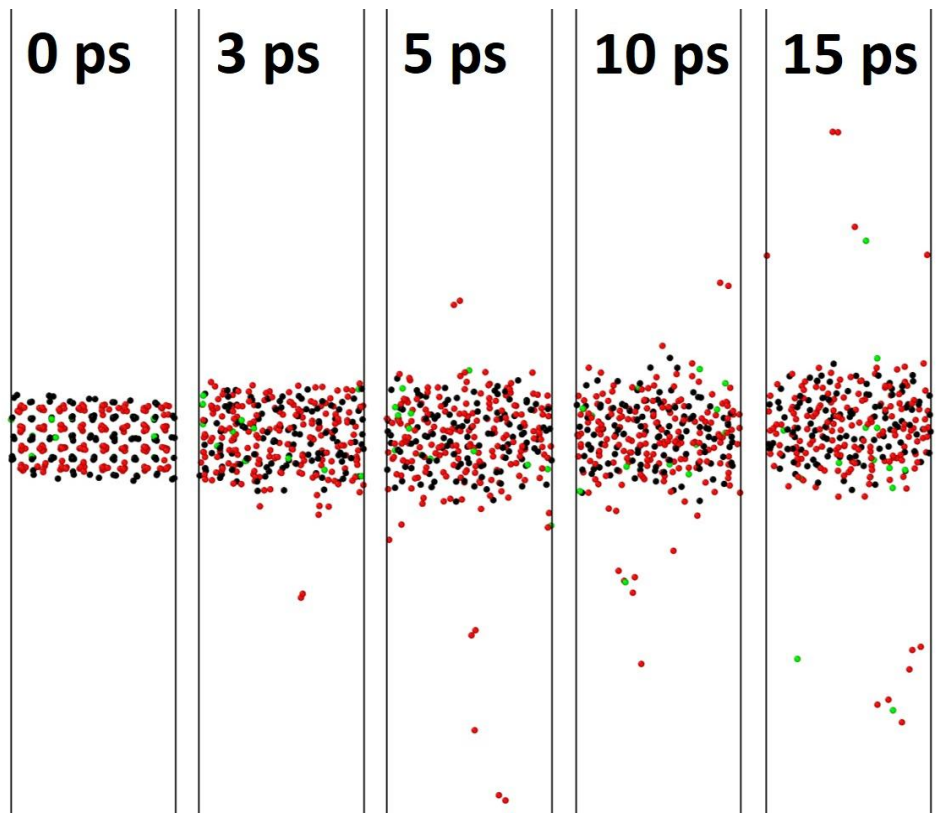

*Figure 18. Atomic snapshots of ITO thin layer irradiated with 1.0 eV/atom dose. Black balls are In; grey balls are Sn; red balls are O.*

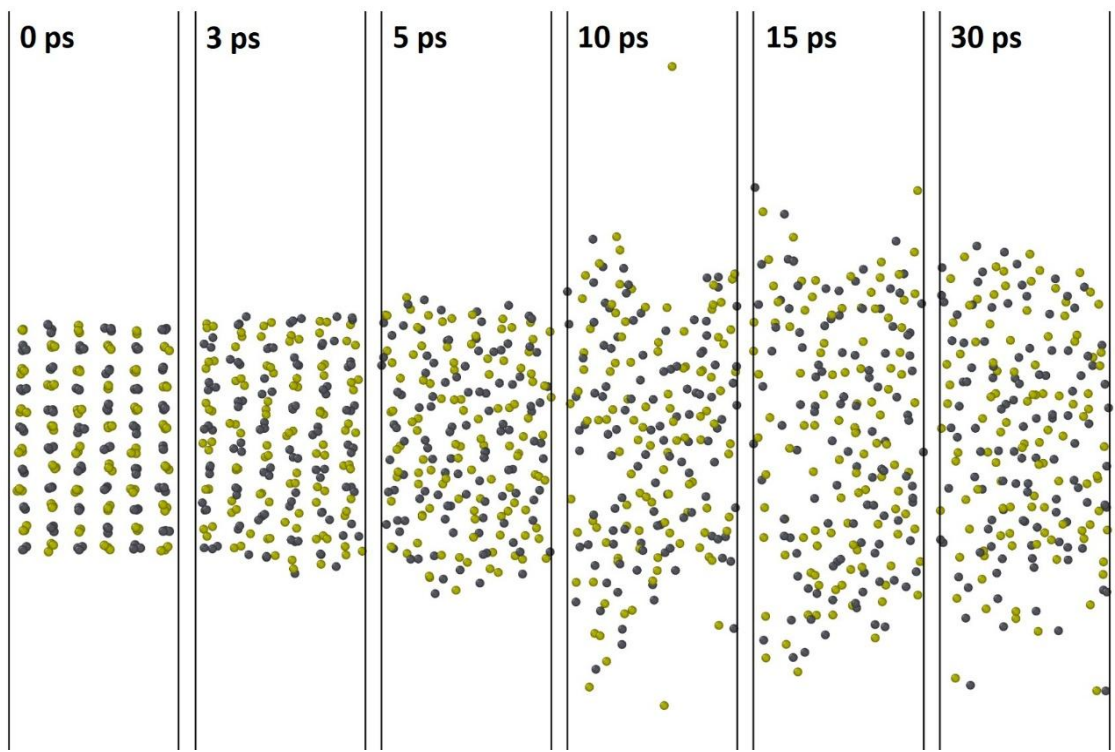

*Figure 19. Atomic snapshots of PbS thin layer irradiated with 0.5 eV/atom dose. Grey balls are Pb; yellow balls are S.*

The band gap in CdTe and ITO thin layers (Figure 20) shrinks to lower values compared to bulk materials irradiated with the same dose (cf. Figure 10). However, the opposite is observed in PbS, which is consistent with the expansion observed in the layer (cf. Figure 19) and the pressure profile calculated for the material (cf. Figure 11), and is in good agreement with the experimental reports on the effects of PbS films thickness on the optical properties of the material[59].

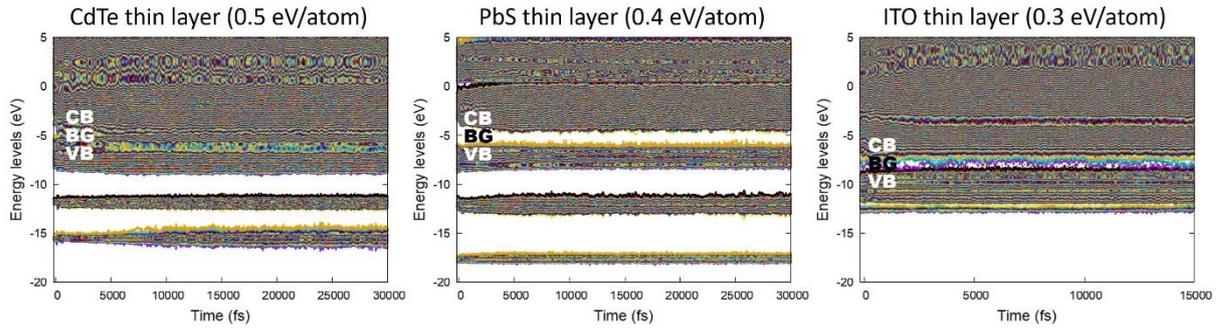

Figure 20. Electronic energy levels (molecular orbitals, band structure) in CdTe, PbS, and ITO thin layers irradiated with different doses. VB, BG, and CB, respectively, mark the valence band, band gap, and conduction band.

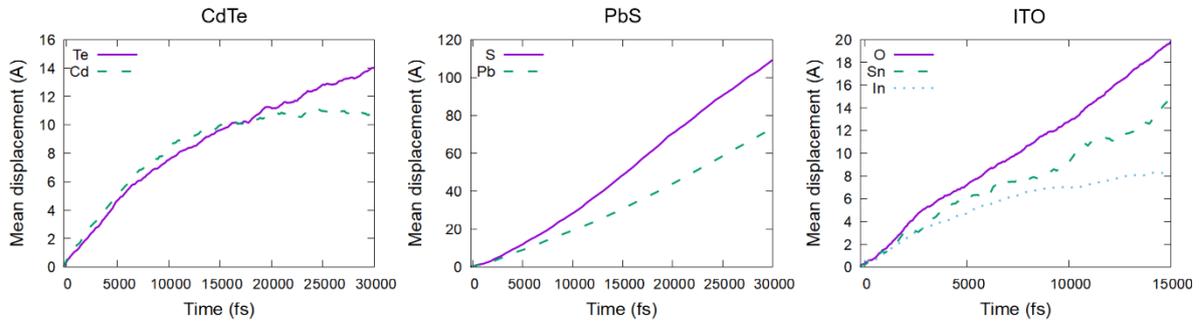

Figure 21. Mean displacement of each species in CdTe, PbS, and ITO thin layers irradiated with 1 eV/atom.

### E.  Damage threshold fluence

Having evaluated the damage threshold doses in the studied materials, they can be converted into the incoming fluence threshold [33]. Such a conversion assumes normal photon incidence, no nonlinear effects, no particle and energy transport in the sample, and no electron or photon emission from the surface. The bulk threshold fluences in CdTe, PbS, ITO, and CdS for comparison, are shown in Figure 22, using EPICS2023 photoabsorption cross sections for the conversion. The damage thresholds are relatively close to one another in all the studied photovoltaic materials, except for different sudden jumps due to different ionization potentials of various shells in different elements. The low damage threshold calculated for ITO is qualitatively supported by the experimental observation in Ref. [60]. These estimates may guide future experiments and application of XUV/X-ray-irradiation of the photovoltaic materials studied.

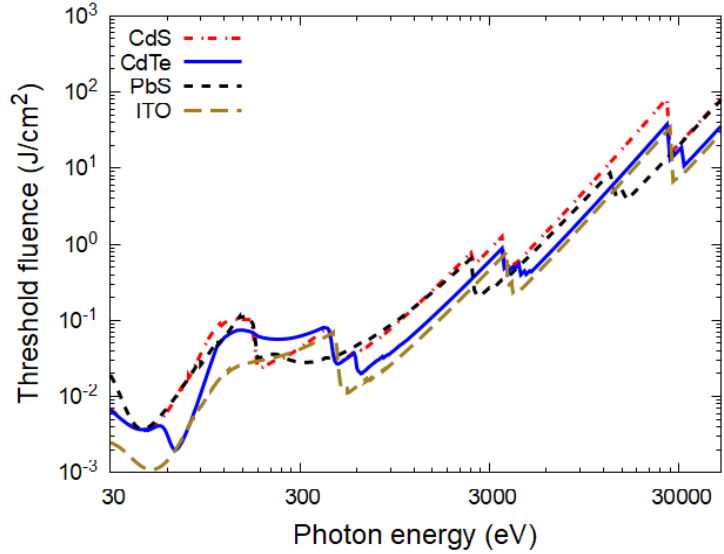

*Figure 22. Damage threshold fluences in CdTe, PbS, ITO, and CdS as functions of the photon energy estimated from the damage doses predicted with XTANT-3.*

## IV. Conclusions

Ultrafast XUV/X-ray irradiation of CdTe, PbS and ITO was modelled with the state-of-the-art hybrid code XTANT-3. CdTe transiently disorders at irradiation doses above ~0.4-0.5 eV per atom, while ITO and PbS disorder at ~0.3-0.4 eV/atom and ~0.2-0.3 eV/atom, respectively. The damage threshold fluence vs. XUV/X-ray photon energy is also estimated for all studied materials (and CdS for comparison).

At the threshold doses, the melting induced is mainly thermal, triggered by the electron–phonon coupling, heating the atomic system. All the materials also exhibit nonthermal melting at higher doses: CdTe at 0.8 eV/atom, PbS at 0.9 eV/atom, and ITO at 1 eV/atom. CdTe and PbS may transiently form semiconducting melted states in the dose intervals between 0.5 and 0.7 eV/atom while turning into metallic liquid at higher doses. CdTe and ITO transiently exhibit superionic states with coexisting solid and liquid sublattices.

The threshold doses increase if energy sinks from the samples and corresponding recrystallization are taken into account. CdTe appears to have the highest recrystallization degree among the studied materials. Below the threshold dose of 1.5 eV/atom, the band gap of each material returns to its original value. With the increase of the dose, the cooled state becomes more amorphous, with correspondingly smaller band gap until an equilibrium value is reached. The results suggest that femtosecond lasers may be useful in tuning the band gap of photovoltaic semiconductors.

At the deposited doses of 0.6 eV/atom in CdTe, and 0.4 eV/atom in PbS and ITO, material ablation from the surface occurs, respectively emitting Te, S, and O/Sn-O aggregates at the characteristic timescale of ~10 ps.

## V. Conflicts of interest

There are no conflicts to declare.

## VI. Data and code availability

The code XTANT-3 used to simulate irradiation effects is available from [30]. Calculated electronic thermal parameters (electronic heat capacity, heat conductivity, and electron-phonon coupling) are publicly available from https://github.com/N-Medvedev/XTANT-3_coupling_data.

## VII. Acknowledgments

We thank L. Juha for fruitful discussions. Computational resources were provided by the e-INFRA CZ project (ID:90254), supported by the Ministry of Education, Youth and Sports of the Czech Republic. NM thanks the financial support from the Czech Ministry of Education, Youth, and Sports (grant nr. LM2023068). The authors gratefully acknowledge the financial support from the European Commission Horizon MSCA-SE Project MAMBA [HORIZON-MSCA-SE-2022 GAN 101131245].